\documentclass[12pt]{article}
\begin{document}
\begin{center}
{\bf{ A Scaling Relation in Inhomogeneous Cosmology with k-essence scalar fields
}}

\bigskip

{\bf{
Debashis Gangopadhyay$^\ast$ and Somnath Mukherjee$^\dagger$

\medskip
{$\ast$Department of Physics, Ramakrishna Mission Vivekananda University, Belur Matth, Howrah-711202, India.\\
debashis@rkmvu.ac.in\\

$^\dagger$Dharsa Mihirlal Khan Institution[H.S],
            P.O:-New G.I.P Colony,  Dist:-Howrah-711112, India.\\
sompresi@gmail.com}}}
\end{center}
 
\begin{center}
Abstract  
\end{center} 

We obtain a scaling relation for spherically symmetric k-essence scalar fields 
$\phi(r,t)$ for an inhomogeneous cosmology with the Lemaitre-Tolman- Bondi (LTB) metric. We show that this scaling relation reduces to  the known relation for  a homogeneous cosmology when the LTB metric reduces to the Friedmann-Lemaitre-Robertson-Walker (FLRW) metric under certain identifications of the metric functions. A k-essence lagrangian is set up and the Euler-Lagrangian equations solved assuming $\phi(r,t)=\phi_{1}(r) + \phi_{2}(t)$. The solutions enable the LBT metric functions to be related to the fields. The  LTB inhomogeneous universe exhibits late time  accelerated expansion i.e.cosmic acceleration driven by negative pressure.

\section{Introduction}

Inhomogeneous cosmological models \cite{R1,R2} are those exact solutions of Einstein's equations that contain at least a subclass of non-vacuum , non-static FLRW solutions as a limit. Inhomogeneous models of the universe arise from perturbations of the FLRW metric. A natural question is whether the phenomenon of late time accelerated expansion  persists in such perturbations of the FLRW metric. One relevant spherically symmetric , inhomogeneous metric is the plane Lemaitre-Bondi-Tolman metric given by 
\begin{equation}
ds^{2}=-dt^{2}+Y'^{2}dr^{2}+Y^{2}[d{\theta}^{2}+\sin^{2}\theta d{\phi}^{2}]
\end{equation}
where $Y=Y(r,t)$ so that $Y'=\frac{\partial Y}{\partial r}$

The stress-energy tensor is \cite{R3}
\begin{equation}
T_{ab}=(\rho+P)u_{a}u_{b}+\left(P-\frac{\Lambda}{8 \pi G}\right)g_{ab}
\end{equation}
where we have included the cosmological constant in Einstein's equations.
Here it should be noted that  \textit {we will not take $\Lambda$ to be the source of 
dark energy}. $k-$essence scalar fields will play that role.
Let 
\begin{equation}
 Y(r,t)=f^{\frac{2}{3}}(r,t)
\end{equation}
Then Einstein's equations (with the cosmological constant $\Lambda$) yield 
\begin{equation}
\rho+\Lambda=\frac{4}{3}\frac{\dot{f}\dot{f}'}{ff'}
\end{equation}
 and
\begin{equation}
P-\Lambda=-\frac{4}{3}\frac{\ddot{f}}{f}
\end{equation}
Here $8\pi G\equiv =1$.
Now let
\begin{equation}
f(r,t)=R(r)T(t)
\end{equation}
 Einstein's equation takes the form of
\begin{equation}
\rho=\frac{4}{3}\left(\frac{\dot{T}}{T}\right)^{2}-\Lambda
\end{equation}
\begin{equation}
P=-\frac{4}{3}\frac{\ddot{T}}{T}+\Lambda 
\end{equation}
From(7)  we get
 \begin{equation}
\dot{\rho}=\frac{8}{3}\frac{\dot{T}}{T}\left[\frac{\ddot{T}}{T}-\left(\frac{\dot{T}}{T}\right)^{2}\right]
\end{equation}
From (8) we get
 \begin{equation}
\frac{\ddot{T}}{T}=-\frac{3}{4}P+\frac{3}{4}\Lambda
\end{equation}
Again from (7) 
\begin{equation}
\left(\frac{\dot{T}}{T}\right)^{2}=\frac{3}{4}\rho+\frac{3}{4}\Lambda 
\end{equation}
Putting (10) and (11) in (9) :
\begin{equation}
\dot{\rho}=\frac{8}{3}\frac{\dot{T}}{T}\left[-\frac{3}{4}P-\frac{3}{4}\rho\right]=-2\frac{\dot{T}}{T}(\rho+P)
\end{equation}
 This gives
\begin{equation}
\dot{\rho}+2\frac{\dot{T}}{T}(\rho+P)=0
\end{equation}
This is the conservation equation for energy.

\section{Scaling relation for the fields }

 The Lagrangian for the $k-$essence field is taken as \cite{R4} 
\begin{equation}
{\mathcal L}= -V(\phi)F(X)
\end{equation}
where
\begin{equation}
 X={{1\over2}{\partial_{\mu}{\phi}{\partial^{\nu}{\phi}}}}=\frac{1}{2}{\dot{\phi}^{2}}-\frac{1}{2}(\nabla\phi)^{2}
\end{equation}
Energy density $\rho$ for k-essence field is given by
\begin{equation}
\rho=V(\phi)[F(X)-2X F_{X}]
\end{equation}
with 
$F_{X}={dF(X)\over{dX}}$.
The pressure $P$ is given by (14)
\begin{equation}
P={\mathcal L}=-V(\phi)F(X)
\end{equation}
Differentiating (16) w.r.t time $t$
\begin{equation}
\dot{\rho}=-[2X F_{X}-F(X)]\dot{\phi}V_{\phi}-\dot{\phi}\ddot{\phi}V(\phi)[2X F_{XX}+F_{X}]
\end{equation}
where $V_{\phi}={dV\over{d\phi}}$, $F_{X}={dF(X)\over{dX}}$ and $F_{XX}={dF_{X}\over{dX}}$ 

Substituting (16),(17) and (18) in (13), the conservation equation  becomes
\begin{equation}
-[2XF_{X}-F(X)]\dot{\phi}V_{\phi}
-\dot{\phi}\ddot{\phi}V(\phi)[2XF_{XX}+F_{X}]
\nonumber\\
+2\frac{\dot{T}}{T}[V(\phi)F(X)-2V(\phi) X F_{X}-V(\phi)F(X)]=0
\end{equation}
On further simplification this becomes
\begin{equation}
-[2X F_{X}-F(X)]\dot{\phi}V_{\phi}-\dot{\phi}\ddot{\phi}V(\phi)[2X F_{XX}+F_{X}]
\nonumber\\
-4\frac{\dot{T}}{T}V(\phi)X F_{X}=0
\end{equation}
i.e.
\begin{equation}
[2X F_{X}-F(X)]\dot{\phi}V_{\phi}+\dot{\phi}\ddot{\phi}V(\phi)[2X F_{XX}+F_{X}]
\nonumber\\
+4\frac{\dot{T}}{T}V(\phi)X F_{X}=0
\end{equation}
i.e.
\begin{equation}
[2X F_{XX}+F_{X}]\dot{\phi}\ddot{\phi}
\nonumber\\
+4\frac{\dot{T}}{T}X F_{X}+[2X F_{X}-F(X)]\dot{\phi}\frac{V_{\phi}}{V(\phi)}=0 
\end{equation}
We take $V(\phi)$ to be constant. Hence (22) becomes
\begin{equation}
[2XF_{XX}+F_{X}]\dot{\phi}\ddot{\phi}+4\frac{\dot{T}}{T}X F_{X}=0
\end{equation}

Now $X={{1\over2}{\partial_{\mu}{\phi}{\partial^{\nu}{\phi}}}}=\frac{1}{2}{\dot{\phi}^{2}}-\frac{1}{2}(\nabla\phi)^{2}$. So $\dot{X}=\dot{\phi}\ddot{\phi}-\frac{1}{2}\frac{d}{dt}[(\nabla\phi)^{2}]$. Thus (23) becomes
\begin{equation}
[2X F_{XX}+F_{X}]\left(\dot{X}+\frac{1}{2}\frac{d}{dt}[(\nabla\phi)^{2}]\right)+4\frac{\dot{T}}{T}X F_{X}=0
\end{equation}
Or
\begin{equation}
[2X F_{XX}+F_{X}]\dot{X}+4\frac{\dot{T}}{T}X F_{X}
\nonumber\\
+[2X F_{XX}
+F_{X}]\frac{1}{2}\frac{d}{dt}[(\nabla\phi)^{2}]=0
\end{equation}
Now let  $\phi=\phi_{1}(r)+\phi_{2}(t)$. Then (25) 
reduces to
\begin{equation}
[2X F_{XX}+F_{X}]\dot{X}+4\frac{\dot{T}}{T}X F_{X}=0
\end{equation}
i.e.
\begin{equation}
[\frac{2}{ F_{X}}\frac{d F_{X}}{dX}+\frac{1}{X}]dX=-4\frac{dT}{T}
\end{equation}
Integrating both sides
\begin{equation}
\int{\frac{dX}{X}}+\int{\frac{2}{F_{X}}d F_{X}}=-4\int{\frac{dT}{T}}
\end{equation}
i.e.
\begin{equation}
\int{\frac{dX}{X}}+2\int{d\ln F_{X}}=-4\int{\frac{dT}{T}}
\end{equation}
i.e.
\begin{equation}
\ln X+2\ln F_{X}=-4\ln T+\ln C
\end{equation}
 where $\ln C$ is an integration constant. So
\begin{equation}
\ln X+\ln {F_{X}}^{2}=\ln CT^{-4}
\end{equation}
Thus we arrive at the scaling relation. 
\begin{equation}
X {F_{X}}^{2}=CT^{-4}
\end{equation}
This is a scaling relation for inhomogeneous cosmology. 
Below we will show that with appropriate identifications,this is the 
same scaling relation present in FLRW homogeneous cosmology with 
dark energy lagrangian defined as in \cite{R4,R5,R6,R7}

\section{ Lagrangian}

Equating k-essence energy density with (7) we get
\begin{equation}
V(\phi)F(X)-2V(\phi)XF_{X}=\frac{4}{3}\left(\frac{\dot{T}}{T}\right)^{2}-\Lambda
\end{equation}
Or
\begin{equation}
XF_{X}=\frac{1}{2}\left[F(X)-\frac{4}{3V(\phi)}\left(\frac{\dot{T}}{T}\right)^{2}+\frac{\Lambda}{V(\phi)}\right]
\end{equation}
From (32) we get
\begin{equation}
F_{X}=\frac{\sqrt{C}}{\sqrt{X}}T^{-2}
\end{equation}
Putting in (34) we get
\begin{equation}
X \frac{\sqrt{C}}{\sqrt{X}}T^{-2}=\frac{1}{2}\left[F(X)-\frac{4}{3V(\phi)}\left(\frac{\dot{T}}{T}\right)^{2}+\frac{\Lambda}{V(\phi)}\right]
\end{equation}
Or
\begin{equation}
F(X)-\frac{4}{3V(\phi)}\left(\frac{\dot{T}}{T}\right)^{2}+\frac{\Lambda}{V(\phi)}=2\sqrt{C}\sqrt{X}T^{-2}
\end{equation}
So that
\begin{equation}
F(X)=+\frac{4}{3V(\phi)}\left(\frac{\dot{T}}{T}\right)^{2}+2\sqrt{C}\sqrt{X}T^{-2}-\frac{\Lambda}{V(\phi)}
\end{equation}
Hence the k-essence lagrangian becomes
\begin{equation}
{\mathcal L}= -V(\phi)F(X)=-\frac{4}{3}\left(\frac{\dot{T}}{T}\right)^{2}
\nonumber\\
-2V(\phi)\sqrt{C}\sqrt{X}T^{-2}+\Lambda 
\end{equation}
Let $q=\ln T$ so that $\frac{\dot{T}}{T}=\dot{q}$.Hence putting in (39) we get
\begin{equation}
{\mathcal L}= -V(\phi)F(X)=-\frac{4}{3}{\dot{q}}^{2}-2V(\phi)\sqrt{C}\sqrt{X}T^{-2}+\Lambda
\end{equation}
So that
\begin{equation}
{\mathcal L}=-c_{1}{\dot{q}}^{2}-c_{2}V(\phi)\sqrt{X}e^{-2q}+\Lambda
\end{equation}
 where $c_{1}=\frac{4}{3}$ and $c_{2}=2\sqrt{C}$.
(41) is the k-essence lagrangian for inhomogeneous scalar field $\phi=\phi(r,t).$

Let $f(\dot{\phi},\nabla\phi)=\sqrt{X}$ so that (41) becomes
\begin{equation}
{\mathcal L}=-c_{1}{\dot{q}}^{2}-c_{2}V(\phi)f(\dot{\phi},\nabla\phi)e^{-2q}+\Lambda
\end{equation}
There are two generalised co-ordinates $q$ and $\phi$.
Euler lagrangian for $q$
\begin{equation}
\frac{\partial{\mathcal L}}{\partial\dot{q}}=-2c_{1}\dot{q}
\end{equation}
and
\begin{equation}
\frac{\partial{\mathcal L}}{\partial q}=2c_{2}V(\phi)f(\dot{\phi},\nabla\phi)e^{-2q}
\end{equation}
therefore 
\begin{equation}
\frac{d}{dt}( \frac{\partial{\mathcal L}}{\partial\dot{q}}) =\frac{\partial{\mathcal L}}{\partial q}
\end{equation}
becomes
\begin{equation}
\frac{d}{dt}(-2c_{1}\dot{q})=2c_{2}V(\phi)f(\dot{\phi},\nabla\phi)e^{-2q}
\end{equation}
or
\begin{equation}
\frac{d}{dt}(\dot{q})=-\frac{c_{2}}{c_{1}}V(\phi)f(\dot{\phi},\nabla\phi)e^{-2q}
\end{equation}
or
\begin{equation}
\frac{d}{dt}(\dot{q})=-\frac{c_{2}}{c_{1}}V(\phi)\sqrt{X}e^{-2q}
\end{equation}
euler lagrangian eqn for $\phi$
From (42)
\begin{equation}
\frac{\partial{\mathcal L}}{\partial\dot{\phi}}=-c_{2}V(\phi)\frac{\partial f}{\partial\dot{\phi}}e^{-2q}
\end{equation}
and
\begin{equation}
\frac{\partial{\mathcal L}}{\partial \phi}=-c_{2}\frac{\partial V}{\partial \phi}f(\dot{\phi},\nabla\phi)e^{-2q}
\end{equation}
Considering constant potential $V(\phi)=constant$,so that $\frac{\partial V}{\partial \phi}=0$,this becomes
\begin{equation}
\frac{\partial{\mathcal L}}{\partial \phi}=0
\end{equation}
Therefore for 
\begin{equation}
\frac{d}{dt}(\frac{\partial{\mathcal L}}{\partial\dot{\phi}})=\frac{\partial{\mathcal L}}{\partial\phi}
\end{equation}
becomes
\begin{equation}
\frac{d}{dt}[-c_{2}V(\phi)\frac{\partial f}{\partial\dot{\phi}}e^{-2q}]=0
\end{equation}
Since $c_{2}$ and $V(\phi)$ are constant hence assuming $c_{2}V(\phi)\neq0$ we get
\begin{equation}
\frac{d}{dt}[\frac{\partial f}{\partial\dot{\phi}}e^{-2q}]=0
\end{equation}
Hence
\begin{equation}
\frac{\partial f}{\partial\dot{\phi}}e^{-2q}=\frac{1}{D}
\end{equation}
where $\frac{1}{D}=constant$.
Now considering  $\phi=\phi_{1}(r)+\phi_{2}(t)$. Then 
\begin{equation}
\frac{\partial f}{\partial\dot{\phi}}=\frac{1}{2\sqrt{X}}\frac{d\phi_{2}(t)}{dt}
\end{equation}
Putting (56) in (55) we get
\begin{equation}
\frac{1}{2\sqrt{X}}\frac{d\phi_{2}(t)}{dt}e^{-2q}=\frac{1}{D}
\end{equation}
Or
\begin{equation}
\sqrt{X}=\frac{D}{2}\frac{d\phi_{2}(t)}{dt}e^{-2q}
\end{equation}
Putting (59) in (48) we get
\begin{equation}
\frac{d}{dt}(\dot{q})=-\frac{c_{2}}{c_{1}}\frac{D}{2}V(\phi)\frac{d\phi_{2}(t)}{dt}e^{-4q}
\end{equation}
Squaring both sides of (58) we get 
\begin{equation}
X=\frac{D^2}{4}\dot{\phi_{2}}^{2}(t)e^{-4q}
\end{equation}
or
\begin{equation}
\frac{1}{2}\dot{\phi_{2}}^{2}(t)-\frac{1}{2}(\nabla{\phi_{1}(r)})^{2}=\frac{D^2}{4}\dot{\phi_{2}}^{2}(t)e^{-4q}
\end{equation}
or
\begin{equation}
\frac{1}{2}\dot{\phi_{2}}^{2}(t)-\frac{D^2}{4}\dot{\phi_{2}}^{2}(t)e^{-4q}=\frac{1}{2}(\nabla{\phi_{1}(r)})^{2}
\end{equation}
Now in (62)the left hand side is a function of time $t$ while the right hand side is a function of $r$ only. Therefore
\begin{equation}
\frac{1}{2}\dot{\phi_{2}}^{2}(t)-\frac{D^2}{4}\dot{\phi_{2}}^{2}(t)e^{-4q}=\frac{1}{2}(\nabla{\phi_{1}(r)})^{2}=B
\end{equation}
where $B$ is a constant. The solutions are
\begin{equation}
\dot{\phi_{2}}(t)^{2}=\frac{2B}{(1-\frac{D^2}{2}e^{-4q})}
\end{equation}
and
\begin{equation}
\phi_{1}(r)=\sqrt{2B} r  
\end{equation}
where we have taken an integration constant to be zero. Substituting (64) in (59) we get
\begin{equation}
\ddot{q}=-\frac{c_{2}}{c_{1}}\frac{D}{2}V(\phi)\frac{\sqrt{2B}}{(1-\frac{D^2}{2}e^{-4q})^{\frac{1}{2}}}e^{-4q}
\end{equation}
Multiplying both sides by $2\dot{q}$ we get
\begin{equation}
2\dot{q}\frac{d\dot{q}}{dt}=-\frac{c_{2}}{c_{1}}\frac{D}{2}V(\phi)\frac{\sqrt{2B}}{(1-\frac{D^2}{2}e^{-4q})^{\frac{1}{2}}}2\frac{dq}{dt}e^{-4q}
\end{equation}
or
\begin{equation}
\frac{d}{dt}{\dot{q}}^{2}=-\frac{c_{2}}{c_{1}}{D}V(\phi)\frac{\sqrt{2B}}{(1-\frac{D^2}{2}e^{-4q})^{\frac{1}{2}}}\frac{dq}{dt}e^{-4q}
\end{equation}
Now
\begin{equation}
\frac{1}{\sqrt{1-\frac{D^2}{2}e^{-4q}}}\frac{dq}{dt}e^{-4q}=\frac{1}{D^{2}}\frac{d}{dt}\left(\sqrt{1-\frac{D^2}{2}e^{-4q}}\right)
\end{equation}
Substituting (69) in (68) we get
\begin{equation}
\frac{d}{dt}{\dot{q}}^{2}=-\frac{c_{2}V(\phi)\sqrt{2B}}{c_{1}D}\frac{d}{dt}\left(\sqrt{1-\frac{D^2}{2}e^{-4q}}\right)
\end{equation}
Solving we get
\begin{equation}
{\dot{q}}^{2}=-\frac{c_{2}V(\phi)\sqrt{2B}}{c_{1}D}\left(\sqrt{1-\frac{D^2}{2}e^{-4q}}\right)
\end{equation}
Here integration constant is chosen as zero.
From (71) we get
\begin{equation}
\dot{q}=\alpha\left[1-\frac{D^2}{2}e^{-4q}\right]^{\frac{1}{4}}
\end{equation}
where $\alpha=\sqrt{-\frac{c_{2}V(\phi)\sqrt{2B}}{c_{1}D}}$. We will always take 
$\alpha$ to be real. (For example, if $c_{1}, c_{2}, D, V$ are all real then taking the negative square root of $2B$ will ensure that $\alpha$ is real etc.)
From (72) we get
\begin{equation}
\int{\left[1-\frac{D^2}{2}e^{-4q}\right]^{-\frac{1}{4}}}dq=\alpha t+\beta
\end{equation}
where $\beta$ is a constant of integration.

Solving we get 
\begin{equation}
t=\frac{1}{\alpha}[-\frac{1}{4}\ln\left(\frac{(1-\frac{D^2}{2}e^{-4q})^{\frac{1}{4}}-1}{(1-\frac{D^2}{2}e^{-4q})^{\frac{1}{4}}+1}\right)
\nonumber\\
-\frac{1}{2}\arctan{(1-\frac{D^2}{2}e^{-4q})^{\frac{1}{4}}}-\beta]
\end{equation}
Using the relations 
\begin{equation}
\ln{\frac{x-1}{x+1}}=-2\coth^{-1}{x}
\end{equation}
eqn (74) becomes
\begin{equation}
t=\frac{1}{2\alpha}[\coth^{-1}{(1-\frac{D^2}{2}e^{-4q})^{\frac{1}{4}}}-\tan^{-1}{(1-\frac{D^2}{2}e^{-4q})^{\frac{1}{4}}}-\beta]
\end{equation}
Since $q=\ln T$ we get
\begin{equation}
t=\frac{1}{2\alpha}[\coth^{-1}{(1-\frac{D^2}{2T^{4}})^{\frac{1}{4}}}-\tan^{-1}{(1-\frac{D^2}{2T^{4}})^{\frac{1}{4}}}-\beta]
\end{equation}
This is the relationship between $t$ and $T$.
Considering the plane Lemaitre-Tolman- Bondi (LTB) metric
\begin{equation}
ds^{2}=-dt^{2}+Y'^{2}dr^{2}+Y^{2}[d{\theta}^{2}+\sin^{2}\theta d{\phi}^{2}]
\end{equation}
It reduces to flat FRW metric in the limit $Y(t,r)\rightarrow a(t)r$ and $Y'(t,r)\rightarrow a(t)$ \cite{R8} :
\begin{equation}
ds^{2}=-dt^{2}+a^{2}(t)[dr^{2}+r^2d{\theta}^{2}+r^2\sin^{2}\theta d{\phi}^{2}]
\end{equation}
Since $f(r,t)=R(r)T(t)$ and $Y=f^{\frac{2}{3}}$ we get
\begin{equation}
Y(r,t)=R^{\frac{2}{3}}(r)T^{\frac{2}{3}}(t)
\end{equation}

\begin{equation}
Y'(r,t)=\frac{\partial Y}{\partial r}
\end{equation}
Thus $Y(t,r)\rightarrow a(t)r$ and  $Y'(t,r)\rightarrow a(t)$ yields 
\begin{equation}
T^{\frac{2}{3}}(t)=a(t)\equiv \bigg(\frac{\dot\phi_{2}^2 D^2}{2\dot\phi_{2}^2 - 4 B}\bigg)^{\frac{1}{6}} 
\end{equation}
where we have used (64).
Again  
\begin{equation}
R^{\frac{2}{3}}(r)=r
\end{equation}
Thus we can say $T^{\frac{2}{3}}(t)$ plays the role of cosmological scale factor for inhomogeneous plane LTB metric. Further (82) means $ T^{4} = a(t)^{6}$. 
So the scaling relation (32) becomes the same as in a homogeneous 
FLRW cosmology, i.e.,  $XF_{X}^{2}=Ca(t)^{-6}$. 

An interesting aspect of these solutions are that \textit{in the limit that the LTB metric reduces to the FLRW metric, the FLRW metric can be written completely in terms of the dark energy scalar fields as follows} :
\begin{equation}
ds^2\nonumber\\
=dt^2-\bigg(\frac{\dot\phi_{2}^2 D^2}{2\dot\phi_{2}^2 - 4 B}\bigg)^{\frac{1}{6}}
2B [ d\phi_{1}^2 + \phi_{1}^2 d\theta^2 + \phi_{1}^2 sin^2\theta d\Phi^2]\nonumber\\
\end{equation}

\section{ Deceleration parameter $Q$}
Deceleration parameter is given by
\begin{equation}
Q=-\frac{\ddot{a}a}{{\dot{a}}^{2}}
\end{equation}
From (82) we get 
\begin{equation}
T(t)=a^{\frac{3}{2}}(t)
\end{equation}
Hence
\begin{equation}
q=\ln T=\ln a^{\frac{3}{2}}=\frac{3}{2}\ln a
\end{equation}
Hence
\begin{equation}
\frac{3}{2}\ln a=q
\end{equation}
Therefore
\begin{equation}
a=e^{\frac{3q}{2}}
\end{equation}
Hence
\begin{equation}
\dot{a}=\frac{2}{3}\dot{q}e^{\frac{3q}{2}}
\end{equation}
and
\begin{equation}
\ddot{a}=\frac{2}{3}\ddot{q}e^{\frac{3q}{2}}+\frac{4}{9}{\dot{q}}^{2}e^{\frac{3q}{2}}
\end{equation}
Putting (89),(90) and (91) in (85) we get
\begin{equation}
Q=-1-\frac{3}{2}\frac{\ddot{q}}{{\dot{q}}^{2}}
\end{equation}
Dividing (66) by (71) we get
\begin{equation}
\frac{\ddot{q}}{{\dot{q}}^{2}}=-1+(1-\frac{D^2}{2}e^{-4q})^{-1}
\end{equation}
Substituting (93) in (92) we get
\begin{equation}
Q=-1-\frac{3}{2}[-1+(1-\frac{D^2}{2}e^{-4q})^{-1}]
\end{equation}
From (76) it is seen that for late time cosmology as $t\rightarrow\infty$,$q\rightarrow\infty$ thus (94) becomes
\begin{equation}
Q=-1
\end{equation} 
which gives an accelerated universe\cite{R9,R10}.

\section{Equation of state parameter$\omega$}
Equation of state parameter is given by 
\begin{equation}
\omega=\frac{P}{\rho}
\end{equation}
Hence from (7) and (8) we get
\begin{equation}
\omega=\frac{P}{\rho}=\frac{-\frac{4}{3}\frac{\ddot{T}}{T}+\Lambda}{\frac{4}{3}\left(\frac{\dot{T}}{T}\right)^{2}-\Lambda}
\end{equation}
Since $q=\ln T$ this becomes
\begin{equation}
\omega=\frac{-\frac{4}{3}(\ddot{q}+{\dot{q}}^{2})+\Lambda}{\frac{4}{3}{\dot{q}}^{2}-\Lambda}
\end{equation}
We are interested in late time cosmologies ,hence considering $q\rightarrow\infty$ for $t\rightarrow\infty$ ,(66) becomes $\ddot{q}\rightarrow0$ and (71) yields ${\dot{q}}^{2}=-\frac{c_{2}V(\phi)\sqrt{2B}}{c_{1}D}={\alpha}^{2}$,Hence putting in (98) equation of state parameter becomes
\begin{equation}
\omega=\frac{-\frac{4}{3}{\alpha}^{2}+\Lambda}{\frac{4}{3}{\alpha}^{2}-\Lambda}=-1
\end{equation}
This satisfies dark energy pressure condition which is negative i.e.,negative pressure generates late time cosmic acceleration.

\section{Conclusion}
 
We consider dark energy in an inhomogeneous universe characterised by the
Lemaitre-Tolman- Bondi metric.  
A scaling relation for spherically symmetric k-essence scalar fields 
$\phi(r,t)$ is obtained which reduces to  the known relation for  a homogeneous cosmology when the LTB metric reduces to the Friedmann-Lemaitre-Robertson-Walker (FLRW) metric under certain identifications of the metric functions. A k-essence lagrangian is set up and the Euler-Lagrangian equations solved assuming $\phi(r,t)=\phi_{1}(r) + \phi_{2}(t)$. The solutions enable the LBT metric functions to be related to the fields. The  LTB inhomogeneous universe exhibits late time  accelerated expansion i.e.cosmic acceleration driven by negative pressure.

\end{document}